# Turbulence and turbulent mixing in natural fluids

Carl H. Gibson, Univ. Cal. San Diego,

Departments of MAE and SIO, Center for Astrophysics and Space Sciences

La Jolla, CA 92097-0411, cgibson@ucsd.edu, USA

**Abstract**

Turbulence and turbulent mixing in natural fluids begins with big bang turbulence powered by spinning combustible combinations of Planck particles and Planck antiparticles. Particle prograde accretions on a spinning pair releases 42% of the particle rest mass energy to produce more fuel for turbulent combustion. Negative viscous stresses and negative turbulence stresses work against gravity, extracting mass-energy and space-time from the vacuum. Turbulence mixes cooling temperatures until strong-force viscous stresses freeze out turbulent mixing patterns as the first fossil turbulence. Cosmic microwave background temperature anisotropies show big bang turbulence fossils along with fossils of weak plasma turbulence triggered as plasma photon-viscous forces permit gravitational fragmentation on supercluster to galaxy mass scales. Turbulent morphologies and viscous-turbulent lengths appear as linear gas-proto-galaxy-clusters in the Hubble ultra-deep-field at z~7. Proto-galaxies fragment into Jeans-mass-clumps of primordial-gas-planets at decoupling: the dark matter of galaxies. Shortly after the plasma to gas transition, planet-mergers produce stars that explode on overfeeding to fertilize and distribute the first life.

1. Introduction

A flood of new observations from space telescopes and automated ground based telescopes in a wide range of frequency bands contradict the standard ΛCDMHC cosmological model [1]. CDM means cold-dark-matter, HC means hierarchical clustering, and Λ is dark energy: all incorrect. The old-cosmology model fails because basic concepts of fluid mechanics are neglected. A new cosmology termed hydro-gravitational-dynamics (HGD) includes turbulence, turbulent mixing, fossil turbulence, intermittency, viscosity, diffusivity, entropy, enthalpy, collisional gas dynamics and the complex nonlinear transport processes of stratified, rotating fluids [2,3]. An explanation of beamed-zombie-turbulence-maser-action (BZTMA) mixing chimneys (generic to natural fluid transport processes) is beyond the scope of the present paper but is required in HGD [3 p49]. Dark energy $\Lambda$ was introduced by Einstein as a necessary antigravity force term in his general relativity theory to prevent the expansion of the universe, but was abandoned by Einstein when observations showed that the universe is expanding.

Old-cosmology claims 75% of the mass-energy of the universe is dark energy. $\Lambda$ is revealed by new-cosmology (HGD) to be a systematic dimming error of supernova brightness observations caused by primordial planets in clumps as the dark matter of galaxies, as illustrated in Fig. 8 ($\Lambda = 0$). Dimming is due to light scattering from evaporated ambient planet atmospheres, not an accelerated expansion of the universe driven by $\Lambda$. These planets in clumps fragmented within proto-galaxies that fragmented along turbulent vortex lines of the plasma epoch just before the transition to gas, Figs. 1 and 2. Such viscosity and turbulence-controlled fragmentations are quite impossible by the standard cosmological model.

Anti-gravity forces are needed to explain the big bang, and are supplied by viscous and turbulence stresses [4,5] that rapidly dissipate. A frictional beginning suggests the fate of the universe is gravitational collapse or a big-bang big-crunch cycle. Viscous stresses are crucial to understanding cosmological phase transitions, where gluon-viscosity powers inflation to terminate the big bang event, plasma photon-viscosity, not CDM, determines the mass and size of the first gravitational structures, and gas-viscosity and Jeans-mass determine the mass of the planets and the mass of their clumps at the plasma to gas transition when galaxy dark matter is formed [6]. Proto-globular-star-clusters PGCs are multi-million solar mass clumps of the primordial-fog-particle PFP earth-mass planets fragmented at decoupling. The galaxy dark matter is detected [7] as "cirrus clouds" of warm "dust" with PGC size clumps in the Milky Way galaxy when viewed by infrared telescopes, or as opaque dense molecular clouds in the optical, Figure 1.

Fig. 1 summarizes infrared and microwave evidence that the Galaxy dark matter is PGC clumps of PFP planets, about $3 \times 10^7$ earth-mass frozen hydrogen-helium planets per galaxy star as predicted from hydro-gravitational-dynamics HGD theory [6] and inferred from quasar microlensing by a galaxy [8]. Schild discovered the twinkling frequency of a galaxy microlensed quasar was not that of stars (years) but of planets (days) as the missing mass. Veneziana et al. 2010 [7 Fig. 2] summarize the 23-3000 Ghz evidence (13 mm to 100 µm) from several platforms of massive "cirrus clouds" of "dust clumps" at all galactic latitudes. "Dust" temperatures modeled to be 7-20 K are significantly warmer than the 2.7 K cosmic background, and match the 15-20 K freezing-boiling point range of HGD heated-cooled frozen hydrogen planets near stars formed by planet mergers.

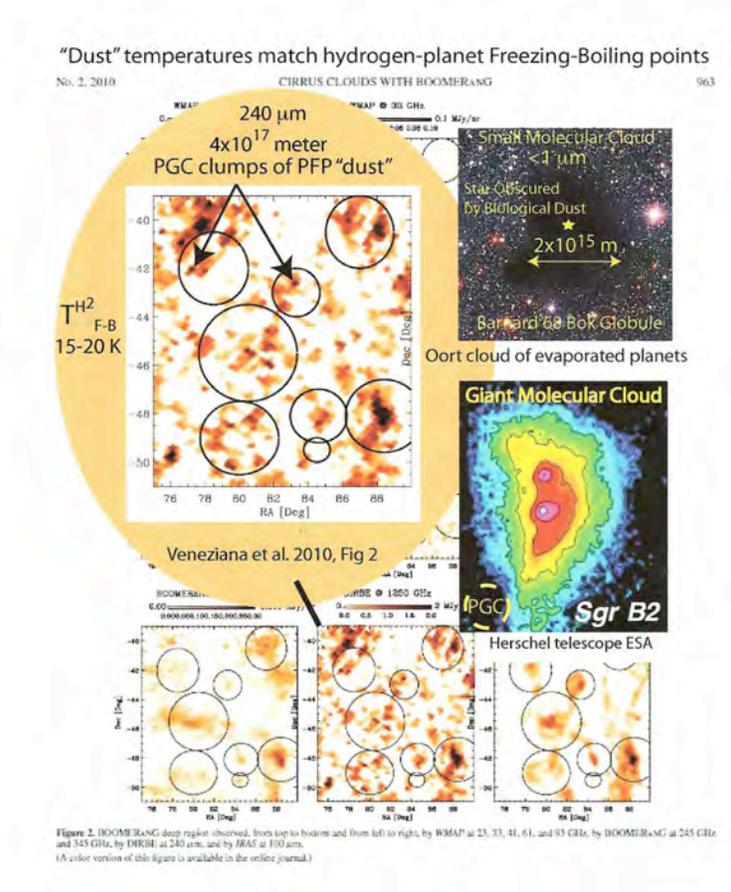

Fig. 1. Infrared and microwave images of the Milky Way halo [7] show 15-20 K clumps of primordial planets comprise the dark matter (missing mass) of the galaxy.

The triple point temperature of hydrogen at pressures expected for an earth-mass planet is ~20K. The planet-clump size ~$4 \times 10^{17}$ m matches that of PGCs from HGD, and the tendency of the clumps to clump suggests the "cirrus dust clouds" are giant molecular clouds like Sgr B2 (Fig. 1 right bottom) formed by sticky clumps of PGCs, where the stickiness results from the large $10^{13}$ m atmospheres formed when frozen-hydrogen-planets evaporate. PFP evaporated planet stickiness is illustrated by small molecular clouds such as Barnard 68 (Fig. 1 right top), where powerful O-B star radiation pressure strips the outer planets surrounding an obscured star, leaving only the $2 \times 10^{15}$ m diameter Bok Globule of the star, here interpreted as an Oort cloud of proto-comet planets.

Giant molecular clouds GMC such as Sgr B2 Fig. 1 (lower right) have long been a mystery, partly because their stars are obscured at optical wavelengths < 1 μm by a complex dust [9, 10] similar to coal dust and oil (Bok Globule upper right). Cirrus dust clouds [7] and GMCs are identical in size and morphology, and cosmological origin. The cirrus "dust" is not powder-like but Earth-mass frozen primordial-gas planets, whose $H^2$ triple point temperatures and pressures match those detected by infrared telescopes. The Jeans mass of PGCs at decoupling is ~ $10^6$ solar masses (~ $10^{36}$ kg). Total PGC mass indicated (~ $10^{42}$ kg) by extrapolating the DIRBE-COBE detections in Fig. 1 (oval insert left) to the full sky significantly exceeds the mass of Milky Way stars.

Cold dark matter hierarchical clustering CDMHC theory makes the highly questionable assumption that a massive population of weakly-collisional nonbaryonic dark matter NBDM particles were somehow created cold enough to condense. Such a material is needed, since the primordial H and $^4$He plasma speed of sound $V_S$ fails the 1902 Jeans length scale $L_J < L_H = ct$ criterion, where $L_J = V_S/(\rho G)^{1/2}$, $c$ is the speed of light, $t$ is the time since the big bang, $L_H$ is the horizon scale or scale of causal connection, $\rho$ is the density and G is Newton's constant of gravity. The first star from CDMHC appears only after a 300 Myr period termed the dark ages (see Fig. 11). Stars, galaxies, and finally superclusters of galaxies form as the CDM halos increase in size in old-cosmology. Voids between superclusters have $10^{24}$ m supercluster size and are full of non-baryonic dark matter (neutrinos) with near-critical flat-universe density ~ $10^{-26}$ kg m$^{-3}$. Observed superclustervoid sizes exceed $10^{25}$ m, with density $< 10^{-35}$ kg m$^{-3}$ [2, 3]. As the proto-superclustervoids of HGD expand from density minimum seeds, weak turbulence is formed that determines the morphology of proto-galaxies fragmented along vortex lines just before decoupling at $10^{13}$ seconds. This model is supported by the recently-repaired Hubble-Space-Telescope ultra-deep-field HUDF observations, shown in Figure 2.

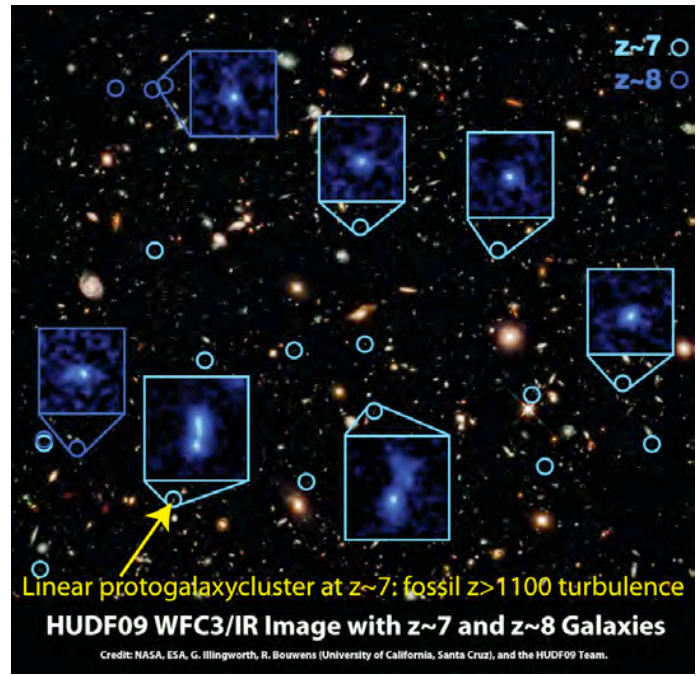

Fig. 2. Hubble Ultra Deep Field (HUDF) images of the dimmest, most distant galaxies and galaxy clusters ever observed, with redshift z~7 and 8. The linear morphology of the protogalaxies shows they were fragmented along plasma vortex lines at z >1100.

Without collisions CDM cannot condense and cannot cluster and will simply diffuse away to $L_{SD} = (D^2/\rho G)^{1/4}$ scales [6], where D is the (very large CDM) diffusivity. No evidence exists of CDM halo seeds condensed in the plasma epoch, or evidence of hierarchical clustering of such seeds required by ΛCDMHC to collect the baryonic matter to make stars, star clusters, galaxies and lastly super-clusters of galaxies.

In the following we review the theory of turbulence and mixing in natural fluids, starting with the turbulent origin of the universe and its evolution to the present. Turbulence and turbulent mixing processes are identical in all natural fluids, whether cosmological, oceanic, atmospheric or laboratory. This important property of turbulence is illustrated by interpreting the remarkable 11 decades of fossil turbulence electron density fluctuations of our local PGC, shown in Fig. 10. The paper is organized into Theory and observations, Observations, Discussion, Conclusions, and a Glossary with Fig. 11 cosmological time-lines and a generic turbulence cascade, Fig. 12.

## 2. Theory and observations

To understand turbulence and turbulent mixing in natural fluids it is first necessary to understand what is turbulence and what is not turbulence but fossil turbulence. Turbulence is defined as an eddy-like state of fluid motion where the inertial-vortex forces of the eddies are larger than any other forces that tend to damp the eddies out [11, 12]. By this definition all irrotational flows are non-turbulent. Because turbulent vorticity is always produced by shear instability at the Kolmogorov scale at a universal critical Reynolds number it is clear that turbulence always starts at small scales and cascades to larger scales, contrary to the well known, but misleading, Richardson 1922 poem ("Big whorls have little whorls that feed on their velocity, etc."). Fossil turbulence is defined as a perturbation in any hydrophysical field produced by turbulence that persists after the fluid is no longer turbulent at the scale of the perturbation. The Reynolds number is the ratio of the inertial vortex force to the turbulence damping viscous force and must exceed a universal critical value for turbulence to exist. The Froude number is the ratio of the inertial vortex force to the turbulence damping buoyancy force, and also must exceed a universal critical value for the flow to be turbulent, by definition. The Rossby number is the ratio of the inertial vortex force to the turbulence damping Coriolis force, etc.

Inertial vortex forces drive the turbulence cascade. Adjacent vortices with the same spin induce **v**x$\omega$ forces towards each, which causes the vortices to merge together into a larger vortex with double the kinetic energy. This is the turbulence energy cascade from small scales to large. Adjacent vortices with opposite spin induce translational and repulsive forces. Figure 3 illustrates the HGD theory of big bang turbulence [5] and big bang fossil turbulence [4] that can occur at Planck temperatures $10^{32}$ K, where Einstein's general relativity theory and quantum mechanics overlap at the Planck length scale $10^{-35}$ m. Vortex lines of secondary turbulence are triggered that wrap around the spinning fireball cylinder, creating adjacent vortices of opposite sign and the anti-gravity forces required to balance Planck gravitational accelerations $\sim 10^{52}$ m s$^{-2}$. Stretching spins up and enlarges the turbulent core and induces a non-turbulent cascade of kinetic energy and fluid volume from the irrotational fluid, as observed in tornadoes, hurricanes, or more easily on the axis of a household blender. Both stretching and entrainment create negative stress and antigravity work against the expansion of space resulting in creation of mass-energy from the vacuum. Mass-energy is created exponentially when the fireball cools to $10^{28}$ K so quark-gluon plasma becomes possible. This strong force phase transition causes inflation from a $10^{-27}$ m fireball to meter size, maintaining the density constant at the Planck value $\sim 10^{97}$ kg m$^{-3}$.

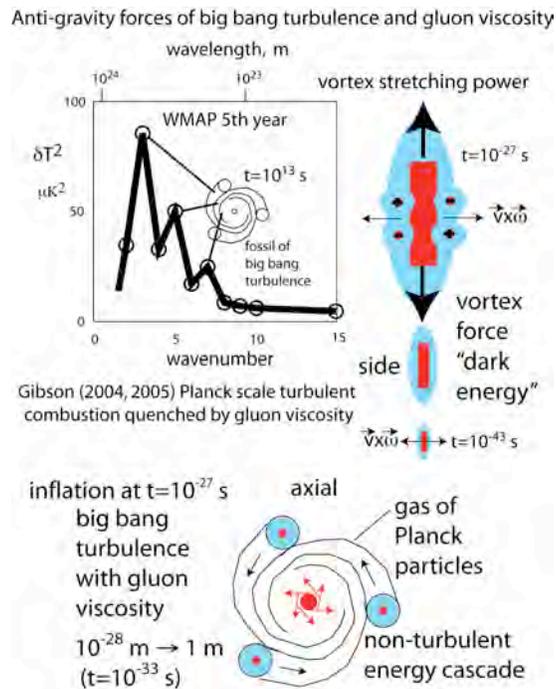

Fig. 3. Anti-gravity forces of big bang turbulent combustion are supplied by turbulence stresses and gluon-viscosity stresses when the fire-ball cools from $10^{32}$ K to $10^{28}$ K.

Because the Planck density is $10^{97}$ kg m$^{-3}$, the mass of the big bang universe within our $ct$ horizon scale $L_H$ is a tiny fraction $10^{-44}$ of the total.

Further evidence of secondary turbulence fossils is shown in Figure 4, from the 5$^{th}$ year data of the WMAP microwave anisotropy background space telescope compared to higher frequency (148 Ghz) ACT radio telescope data [13]. From HGD the first gravitational structures to develop during the plasma epoch are protosuperclustervoids, where density minima expand gravitationally when the horizon scale $ct$ first exceeds the Schwarz viscous-gravitational scale at time $10^{12}$ s. Such density minima are rarefaction waves that expand at the plasma speed of sound $c/3^{1/2}$, where c is the speed of light, for the $10^{13}$ s duration of the plasma epoch. Weak turbulence is created at the void boundaries by spinning up residual big bang vorticity with the "axis of evil" direction [14]. The sonic peak reflects the supervoid size, and the two bumps in the CMB spectrum reflect secondary turbulence vortices.

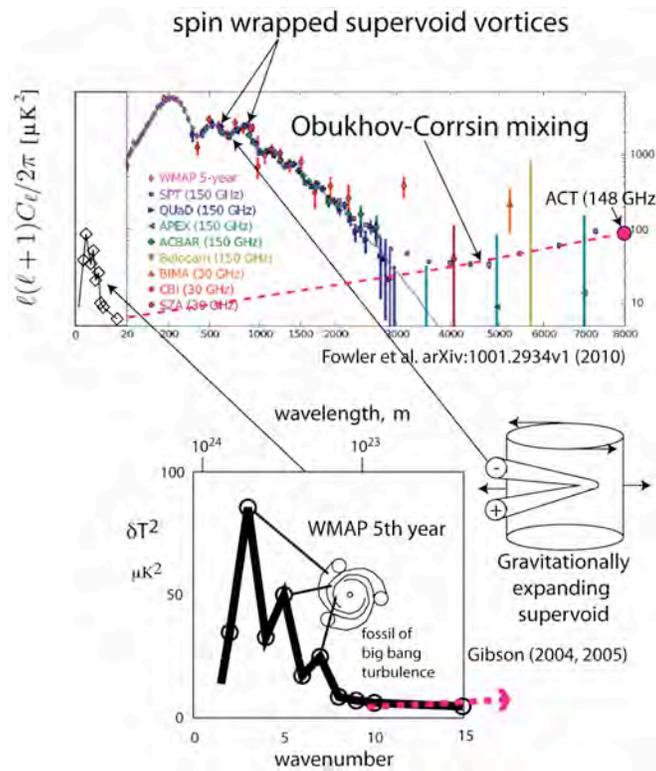

Fig. 4. CMB anisotropy spectra and radio telescope data [13]. Sonic peaks are interpreted as protosuperclustervoids expanding gravitationally at sonic speeds from $10^{12}$ seconds till photon decoupling at $10^{13}$ s (300,000 years) with secondary turbulent vortices. The dashed arrow compares the ACT and WMAP data to an Obukhov-Corrsin fossil big bang turbulent mixing spectrum.

As shown in Fig. 4, the ACT and some other radio telescope spectra are significantly above the ΛCDMHC prediction at wavenumbers in the range 3000-8000, supporting a fit to the Obukov-Corrsin turbulent mixing form expected from HGD.

Figure 5 shows the usual semi-linear presentation of CMB data [15] compared to the relatively low spectral levels of big bang turbulence fossils on such plots, emphasizing the sonic peak and secondary (turbulence) subpeaks.

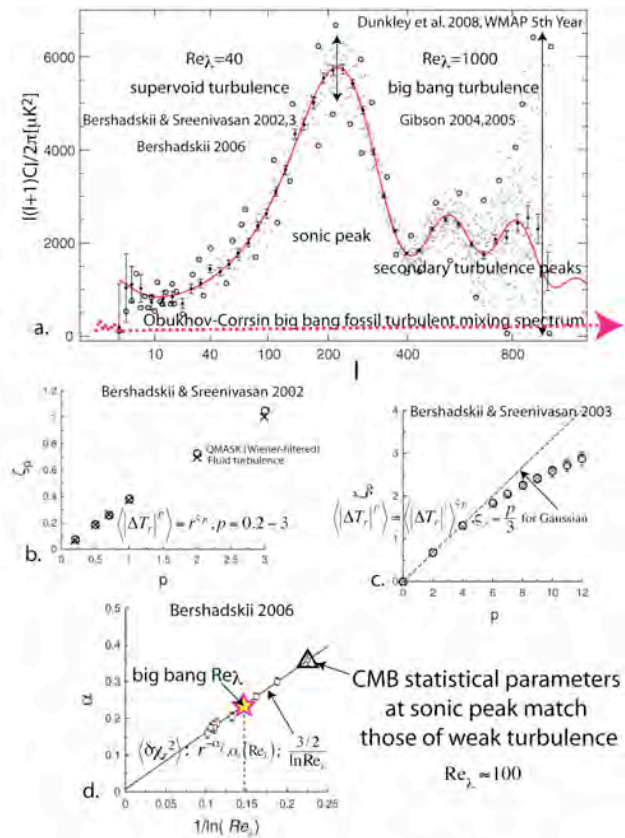

Fig. 5. Fifth year CMB spectra [15] compared to turbulence and turbulent mixing interpretations of Bershadskii & Sreenivasan [16, 17] and Bershadskii [18].

The sonic and subpeaks of Fig. 5a are taken by ΛCDMHC as evidence of plasma sonic oscillations within small mass (fluid mechanically impossible) CDM halos that collect progressively larger masses of plasma and then gas as the halos merge. Here the secondary peaks are interpreted as fossils of secondary turbulence vortices induced by gravitationally expanding supervoids concentrating remnant big bang vorticity of the surrounding protosupercluster plasma. Bershadskii and Sreenivasan [16,17,18] compare statistical parameters of turbulence flows with those of CMB temperature anisotropies near the sonic peak where the signal to noise ratio is high. The correlation is very high, and suggests a small turbulence Taylor microscale Reynolds number $Re_\lambda$ of 40, much smaller than the $Re_\lambda$ of 100 predicted by big bang turbulence theory but significantly above transition values of 5-10. Fossils of big bang turbulence dominate plasma turbulence fossils of the strong force freeze-out at small wavenumbers 2-15 and the highest wavenumbers 3000-8000 detected by radio telescopes. Vertical arrows in Fig. 5 show an increase in spectral intermittency with wave-number that can be attributed to the larger $Re_\lambda$ values of big bang turbulence.

Figure 6abcd summarizes the HGD theory of gravitational structure formation.

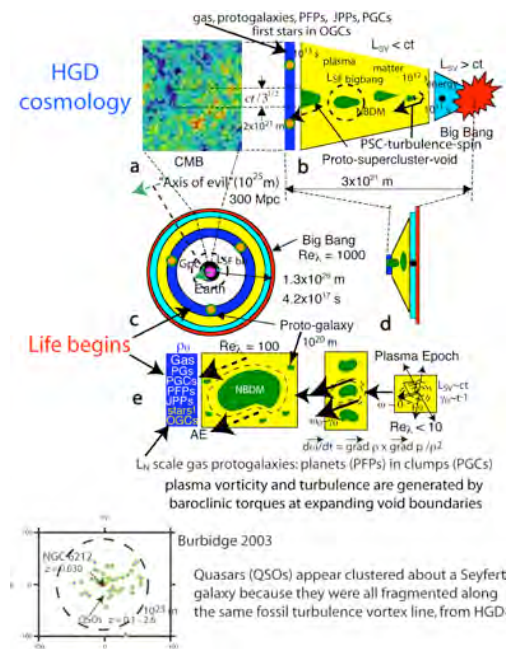

Fig. 6. Summary of HGD cosmology viewing the CBM temperature anisotropies from the earth. Looking out in space is looking back in time with the big bang as the outer shell at 13.7 Gyr. Gravitational structure begins by viscous-gravitational fragmentation of the plasma when $L_{SV}=(\gamma\nu/\rho G)^{1/2}< ct$. Life appears [9,10] at $\sim t<10^{14}$ s, soon after the explosion of the first star to provide the necessary star dust fertilizer. Evidence of galaxies fragmented along fossil plasma turbulence vortex lines is provided (bottom) by clustering of quasars about radio bright (Seyfert) galaxies [19].

Fig. 6abc shows the view of the universe from earth according to general relativity theory, where all observers look back in time as they look out in space to their horizon distance $L_H = ct$ as though each is at the center of the universe. Because the Schwarz viscous-gravitational scale $L_{SV}=(\gamma\nu/\rho G)^{1/2}>L_H$ before $10^{12}$ s, no structure can form, where γ is the rate-of-strain, ν is the kinematic viscosity, ρ is density and G is Newton's gravitational constant. Non-turbulent fragmentation at density minima amplifies fossil big bang vorticity turbulence at protosupercluster PSC scales along the "Axis of Evil" direction detected in the CMB for all low wavenumber spherical harmonics, as well as in the spin alignments and handedness of our nearby galaxies [14], Fig. 6de. The insert at the bottom of Fig. 6 shows a strong clustering of quasars observed about an axial-view Seyfert galaxy [19]. Such clusterings are easier to explain from HGD as fragmentations of protogalaxies along the same fossil turbulence vortex line at the end of the plasma epoch, rather than as ejection of the quasi-stellar-object QSOs from NGC6212 with large intrinsic redshift. Maximum γ values occur on turbulent vortex lines.

The smallest mass objects formed in the plasma epoch are protogalaxies, with Kolmogorov and Obukhov length scales determined by the universal similarity laws of stratified turbulence and the Nomura scale $L_N = 10^{20}$ m and Nomura morphology, where fragmentation is triggered by γ on turbulent vortex lines and on spiraling turbulent fluid particles spinning up and flattening at the base of the vortices. As shown in Fig. 2, galaxies are observed in linear clusters < 1 Gyr after the big bang.

3. **Observations**

An important prediction of HGD cosmology is that viscosity and turbulence govern the formation of gravitational structures. At the end of the plasma epoch the kinematic viscosity decreases from ν values $\sim 10^{26}$ m$^2$ s$^{-1}$ to values of hot primordial gas of only $\sim 10^{13}$ m$^2$ s$^{-1}$. The viscous gravitational fragmentation scale $L_{SV}$ therefore decreases from protogalaxy mass scales to earth mass scales. A strong mismatch develops between the sound speed and light speed in the gas, so fragmentation occurs simultaneously at the Jeans acoustic scale $L_J = V_S \tau_g$, where $V_S$ is the sound speed and $\tau_g$ is the gravitational free fall time $(\rho G)^{-1/2}$. All protogalaxies thus fragmented into protoglobularstarcluster

PGC clumps of primordial fog particle PFP clouds now preserved as the galaxy dark matter. Because the motions were extremely gentle at this time of decoupling, the small long lived stars of old globular clusters appeared in PGCs near the protogalaxy centers, and some of these exploded to fertilize the first life in the many relatively warm primordial soup domains then existing. Evidence of this scenario is emerging from telescopes sensitive in the infrared, as shown in Figure 7 for the Helix [20].

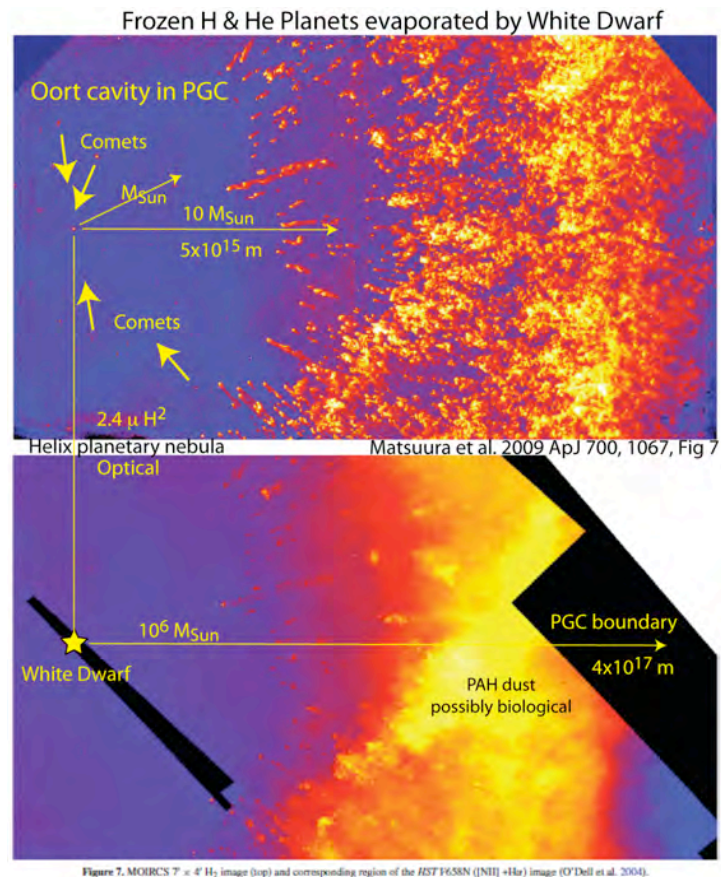

Fig.7. Infrared and optical observations of evaporated Jovian PFP planets JPPs surrounding the dying White Dwarf star of the Helix Planetary Nebula [20].

The images of Helix shown in Fig. 7 compare the infrared 2.4 μ $H^2$ line at the top to the Hubble-Space-Telescope HST optical images [21] at the bottom. Plasma jet radiation from the central White Dwarf evaporates PFP and JPP frozen gas planets at the inner boundaries of the Oort cavity left in the PGC by the star formation and enlarges the cavity by radiation pressure. PAH dust obscures at optical frequencies (bottom) many planets clearly shown in the infrared (top). Proto-comet planets are detected in the infrared falling toward accretion. Presumably their PAH content is smaller than the Oort cavity boundary planets that are accelerated more strongly outward by radiation as they evaporate to larger scales.

Figure 8 gives an HST optical view [21] of the Helix PNe, suggesting (top) that the Supernova Ia dimness evidence of dark energy $\Lambda$ [22] is simply a systematic dimming error of PFP planets as the dark matter of galaxies, depending on the line of sight.

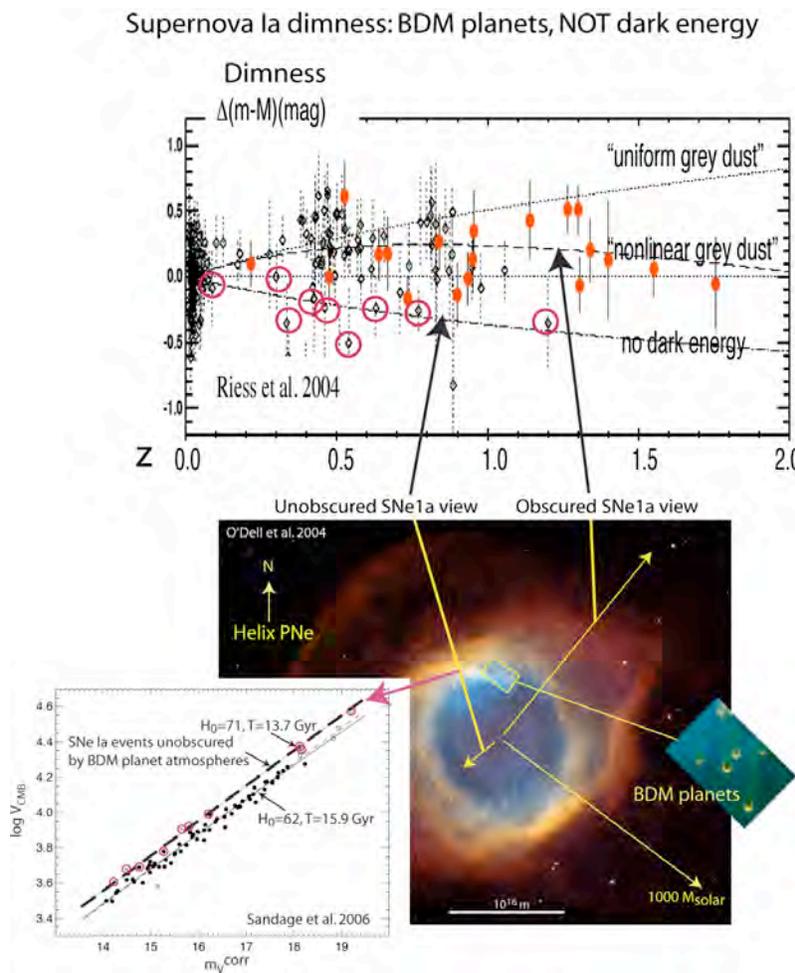

Fig. 8. Observations that show the anomalous dimness of supernova Ia events [22] and anomalously low Hubble constants [23] can be attributed to baryonic dark matter (BDM) planet atmospheres, not dark energy [24]. The nearby Helix planetary nebula PNe at $6.7 \times 10^{18}$ m has a central white dwarf with polar jet that evaporates ambient BDM planets of its PGC. A close-up view is shown in the insert on the right [21].

A similar conclusion is reached about the Sandage et al. 2006 [23] estimates of the Hubble constant $H_0$, indicating the age of the universe is 15.9 Gyr, much larger than the 13.7 Gyr estimate from CMB evidence and HGD theory. Red circles on the upper and lower plots emphasize the brightest observations of SNIa events, interpreted as lines of sight unobscured by atmospheres of evaporated dark matter planets. Dimmer events are interpreted as due to scattering by fossil turbulent density fluctuations in the intercepted evaporated planet atmospheres, rather than increased distance due to anti-gravity accelerations by dark energy $\Lambda$.

Figure 9 shows a Herschel space telescope infrared image of the Eagle Nebula.

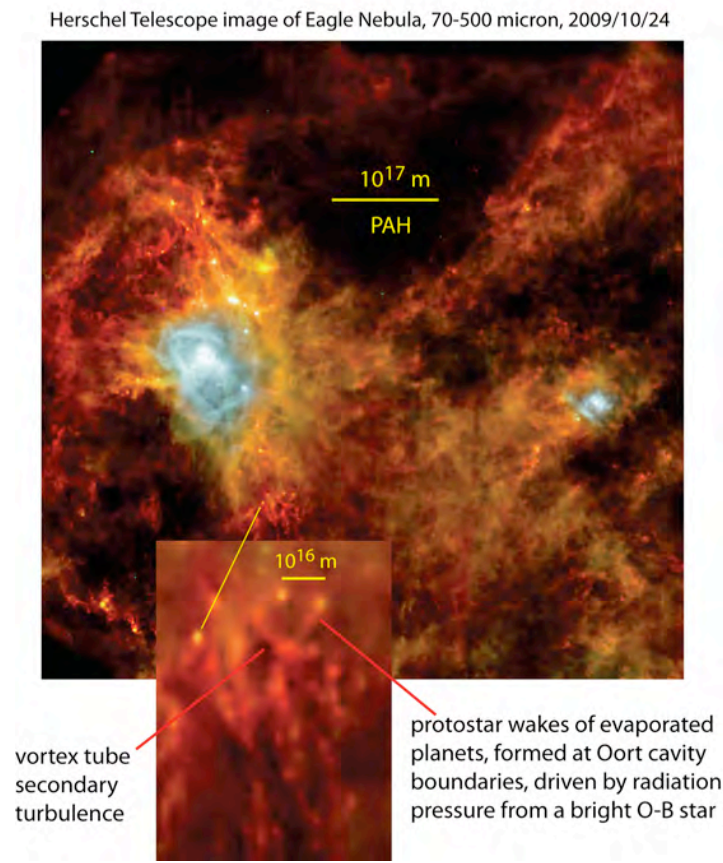

Fig.9. Herschel space telescope images of the Milky Way disk star forming region Eagle Nebula in the infrared bandwidth 70-500 μ. Evidence of disk turbulence is shown by the reduced size of the PGC clumps of PFP planets made dark by PAH dust, and by the protostar wakes of planets heated and evaporated by nearby O-B stars.

Eagle is embedded in the Milk Way spiral galaxy disk, interpreted as a PGC accretion disk in HGD where turbulence and large star formation is enhanced. In Fig. 9 we see protostars in formation surrounded by evaporated planets that provide flow visualization. Powerful radiation for O-B stars evaporate and erode dark matter planets of the protostars, revealing wakes and evidence of secondary turbulence eddies in the bottom insert.

Figure 10 is termed the great power law on the sky by the radio telescope literature, and has been a mystery for nearly 30 years since its $P_{3N} \sim q^{-11/3}$ spectral power law for electron density is identified with that of Kolmogorovian turbulence in the inertial range [25, 26].

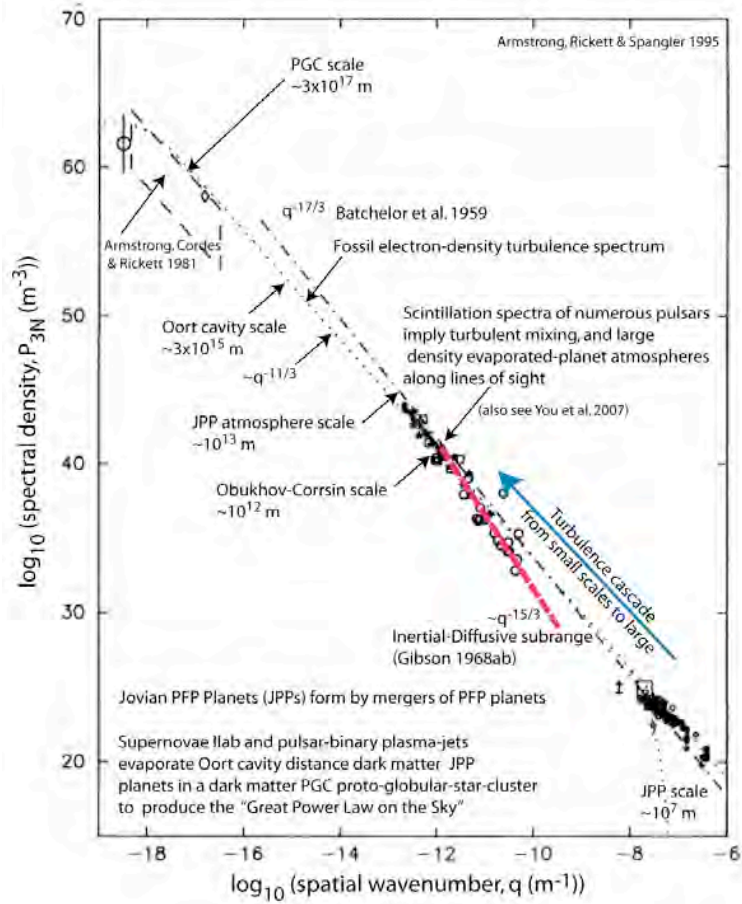

Fig.10. Great power law on the sky. A variety of radio telescope fossil turbulence electron-density $P_{3N}$ power spectra $\sim q^{-11/3}$ are combined over 11 wavenumber q decades suggesting a turbulent mixing origin of supernova driven plasma within the PGC planet clump surrounding the Earth.

The measured spectra of Fig. 10 can be interpreted from HGD cosmology and from turbulence and turbulent mixing theory [11]. Radio telescopes respond to fluctuations of electron density, so this is a fossil Obukhov-Corrsin spectrum of turbulent mixing where electron density is a strongly diffusive passive scalar property for length scales larger than the Obukhov-Corrsin scale [27, 28]. The source of turbulent electron density at wavelength $q^{-1} \sim 10^7$ m is JPP (merged PFP) planets evaporated by supernova events. The red dashed line $P_{3N} \sim q^{-15/3}$ from pulsar scintillation spectra supports the Gibson 1968ab theoretical prediction for a fossil turbulent JPP atmosphere with density $\rho > 10^{-11}$ kg m$^{-3}$, scale $\sim 10^{13}$ m and electron density diffusivity $D \gg \nu$ rather than the Batchelor et al. 1959 $P_{3N} \sim q^{-17/3}$ spectrum shown for comparison. Without the dense planet atmosphere viscous forces would prevent turbulence, so this is further evidence of frozen hydrogen dark matter planets. The cascade to PGC scales $4 \times 10^{17}$ m implies that supernova blasts extend over the full extent of the PGC. Clearly turbulence and turbulent mixing of electron density cascades eleven decades from small JPP scales to large PGC scales as shown by the blue arrow, and then fossilizes. A generic example of the turbulence "inverse" cascade mechanism is described in Fig. 12 of the Glossary.

4. **Discussion**

Evidence of turbulence and turbulent mixing in cosmology and astrophysics has been presented that strongly supports hydrogravitational dynamics HGD cosmological theory rather than the standard collisionless ideal flow model (ΛCDMHC) based on cold-dark-matter and a permanent anti-gravity dark energy open universe Λ. Gluon-viscosity negative stresses terminate the big bang turbulence event and work against the expansion of the universe to produce exponential inflation of mass-energy extracted from the vacuum, Fig. 3. Photon-viscous stresses prevent gravitational structures from forming on fossil big bang turbulence density fluctuations until $10^{12}$ s (30,000 years) after the big bang, and govern a large (supercluster) to small (galaxy) mass fragmentation cascade in the plasma epoch until the plasma to gas transition to protogalaxies fragmented along turbulent vortex lines, supported by observations in Figs. 2-6.

Evidence of turbulence and mixing in astrophysics is provided by radiowave, microwave, infrared and optical space telescope data [7] supporting the HGD prediction that protogalaxies at plasma-gas transition fragment into Jeans mass clumps of earth mass planets destined to be the dark matter of galaxies, Figs. 1, 7-10.

Infrared telescopes Planck and Herschel were launched simultaneously May 14, 2009, so data presently available are in early release form. Many more valuable comparisons with cosmological models should be forthcoming in the future, as well as further astrophysical tests of turbulence and turbulent mixing theories.

5. **Conclusions**

As discussed, many recent observations show the standard ΛCDMHC cosmological model of cosmology must be strongly modified to take basic fluid mechanics into account. Hydrogravitational dynamics HGD cosmology explains the big bang as a Planck scale turbulent combustion event terminated by anti-gravity gluon-viscous-stresses that drive exponential inflation of mass-energy to length scales $\sim 10^{15}$ larger than our present horizon $L_H = ct$. Fluid mechanical negative stresses (anti-gravity dark energies) of the big bang are frictional and temporary, suggesting a closed universe and a series of big-bang big-crunch events. Fossil turbulence CMB patterns of big bang turbulence and turbulence produced by structure formation in the plasma epoch agree precisely with those of terrestrial turbulence both measured and simulated [16-18] in Fig. 5. Universal similarity of turbulence, turbulent mixing, and fossil turbulence based on a revised definition of turbulence [9-12] is also strongly supported by radio telescope detections of local electron density fluctuations (Fig. 10).

Evidence continues to mount that the dark matter of galaxies is frozen hydrogen-helium planets PFPs in proto-globular-star-cluster clumps PGCs, a trillion Earth-mass planets per clump (Figs. 1-9) as first predicted [6] and first observed [8]. All stars are formed by mergers of these primordial planets according to HGD fluid mechanical modifications of the ΛCDMHC standard theory (Fig. 6), and the stars die by supernovae if their accretion of planets fails to cease. Galaxies fragment along turbulent vortex lines (Figs. 2, 6).

Turbulence and turbulent mixing of natural fluids of oceanography, atmospheric science, and engineering practice all depend on recognizing a definition of turbulence based on inertial-vortex forces where irrotational flows are non-turbulent, and where turbulence always cascades from small scales to large [11, 12]. Fossil turbulence patterns (Fig. 10) are ambiguous without this definition [2-6].

Ignorance of stratified turbulence and fossil turbulence behaviors and the effects of extreme intermittency in oceanography and atmospheric sciences can be extremely expensive, not only in cost but in lives. Oil droplet sizes and buoyant plume trapping depths and layer thicknesses are examples of fossil turbulence parameters apparently unexploited by oceanographers and government regulators of petroleum volcanoes. Large undersampling errors are typical of oceanographic and atmospheric turbulence microstructure research from neglect of the extreme intermittency of turbulence and turbulent mixing in natural fluids. Clear air turbulence risks to aircraft flying in thunderstorm regions increase by factors of order $10^6$ at equatorial latitudes where the range of scales of the turbulence cascade is very weakly constrained by Coriolis forces.

6. **Glossary and time-lines**

Many terms and units of fluid mechanics are not familiar to astrophysicists just as many astrophysical terms and units are not familiar to fluid dynamicists. Following is a brief attempt to supply some tools of translation. Distances to astrophysical objects are often not given because until recently these distances were quite uncertain. If given, the normal unit is the parsec, where 1 pc = $3.1 \times 10^{16}$ meters, or the astronomical unit distance from the Earth to the Sun, where 1 AU = $1.5 \times 10^{11}$ meters. Masses are given in solar masses, where 1 $M_{Sun}$ = $2.0 \times 10^{30}$ kg. Time is given in Gyr, where the age of the universe since the big bang is 13.7 Gyr = $4.3 \times 10^{17}$ seconds. Densities range from the Planck density $5.16 \times 10^{96}$ kg m$^{-3}$ of the big bang to a minimum of about $10^{-35}$ kg m$^{-3}$ for present-day galaxysuperclustervoids that began forming at $10^{12}$ seconds (30,000 years) as the first gravitational structures of the plasma epoch. The baryonic (proton-neutron) density at the time of first structure $10^{12}$ seconds was $4 \times 10^{-17}$ kg m$^{-3}$ and is preserved as the density of old globular star clusters OGCs, protoglobularstarclusters PGCs and young globular clusters YGCs. For comparison, the density of a galaxy is about $10^{-21}$ kg m$^{-3}$ and the critical density of a flat universe is $10^{-26}$ kg m$^{-3}$. The size of protogalaxies and galaxy cores is the Nomura scale $10^{20}$ m. The trillion planets per PGC began freezing at about $10^{16}$ seconds (300 Myr) and diffusing out of the core to form the spherical galaxy halo, with diameter ~ $10^{22}$ m and an accretion disk of thickness ~ $10^{19}$ m. Spiral galaxies show star trails of YGCs in the halo triggered by tidal forces of PGCs as they accrete back into the core on the disk [2, 3]. The non-baryonic dark matter (neutrinos) is relatively collisionless (nucleon collision cross section $\sigma$ ~ $10^{-40}$ m$^2$) and diffuses to such large $L_{SD}$ scales >$10^{23}$ m that the galaxy central density is not affected.

Figure 11 contrasts time-lines of HGD and ΛCDMHC cosmologies.

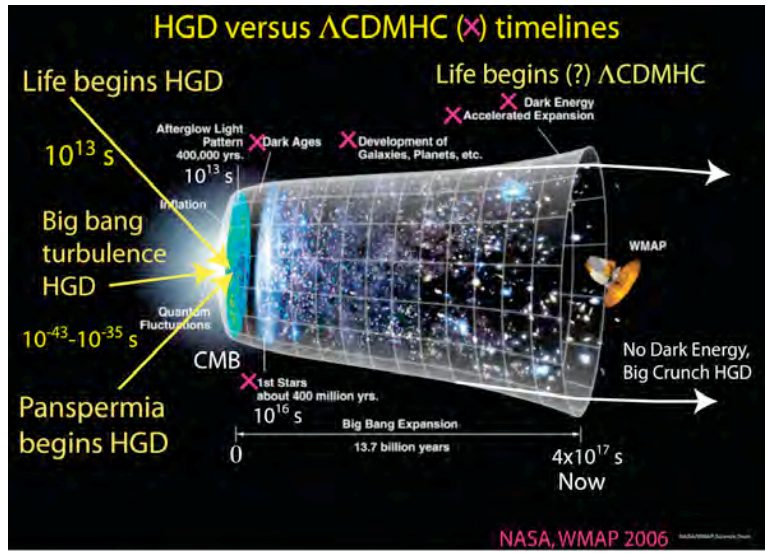

Fig. 11. New-cosmology (HGD) versus old-cosmology (ΛCDMHC) timelines [29].

In Fig. 11 we see time-lines of the two cosmologies. Both begin with a big bang at time zero under Planck conditions, but the anti-gravity effects of turbulence stresses and gluon-viscosity stresses (HGD dark energy) are highly frictional and temporary ($\Lambda = 0$). The entropy produced by inflation to mass-energy ~ $10^{97}$ kg at meter sizes is small due to the high temperature, but implies an eventual gravitational collapse of the universe (big crunch) at some future time. Planck temperatures produced by the big crunch make another big bang probable, with 50% chance of anti-matter dominance rather than matter if the first prograde capture is a Planck particle on a spinning Planck particle pair rather than a Planck anti-particle (giving matter) to trigger big bang turbulent combustion.

Figure 12 demonstrates the generic cascade of turbulence from small scales to large driven by the inertial vortex forces. The formation of a turbulent boundary layer is chosen as a familiar example representative of how turbulence is always created.

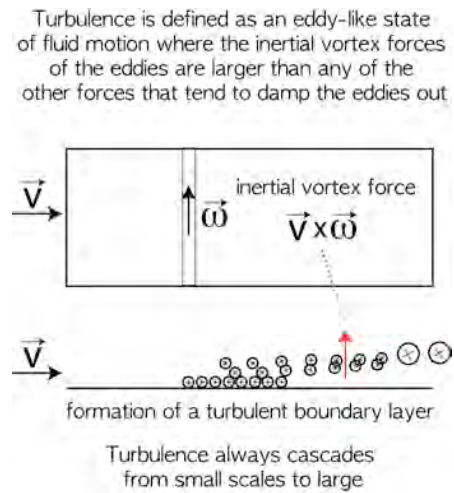

Fig. 12. The definition of turbulence is given based on the inertial vortex force. The formation of a turbulent boundary layer is a generic example.

As shown in Fig. 12, nearby eddies with the same sign are driven together and merge due to inertial vortex forces. This is the basis for Kolmogorovian universal similarity hypotheses [11,12]. Fossil turbulence and fossil turbulence waves cannot be defined without this definition. Fossil turbulence is defined as any perturbation in a hydrophysical field caused by turbulence that is no long turbulent at the scale of the perturbation. If the fluid of Fig. 12 is stably stratified as for the ocean or atmosphere, the turbulence fossilizes when the inertial vortex forces can no longer overturn in the vertical direction at a critical Froude number, so that all the turbulent kinetic energy at the largest scales is converted to fossil vorticity turbulence waves that radiate into the interior of the fluid, providing the dominant vertical mixing process for the ocean and atmosphere and the dominant radial mixing process for stars and other self-gravitating astrophysical objects [2,3]. The accretional cascade of primordial planets to form stars leads to a similar intermittent lognormal distributions of mass density and consequent vast undersampling errors and false exclusion curves by the MACHO, OGLE and EROS microlensing collaborations attempting to sample planetary mass objects as the dark matter of galaxies [29]. Oceanographic textbooks typically start with highly misleading discussions of turbulence that misunderstand the generation mechanism shown in Fig. 12 and falsely claim the turbulence energy cascade is from large scales to small [30].